\documentclass[12pt,a4paper]{article}
\pdfoutput=1

\usepackage{ifthen} 
\newboolean{pdflatex}
\setboolean{pdflatex}{true}

\newboolean{articletitles}
\setboolean{articletitles}{true} 

\newboolean{uprightparticles}
\setboolean{uprightparticles}{false} 

\newboolean{inbibliography}
\setboolean{inbibliography}{false} 

\def\paperauthors{LHCb collaboration}
\def\paperasciititle{Search for the lepton-flavour violating decays B_d,s to e mu} 
\def\papertitle{Search for the lepton-flavour violating decays \Bdsemu} 
\def\paperkeywords{{JHEP}, {LHCb}} 
\def\papercopyright{CERN on behalf of the LHCb collaboration}
\def\paperlicence{CC-BY-4.0}
\def\paperlicenceurl{https://creativecommons.org/licenses/by/4.0/}

\usepackage[top=1in, bottom=1.25in, left=1in, right=1in]{geometry}

\usepackage{microtype}
\usepackage{lineno}  
\usepackage{xspace} 
\usepackage{caption} 
\usepackage{url}

\usepackage{graphicx}  
\usepackage{color}
\usepackage{colortbl}
\graphicspath{{./figs/}} 

\usepackage{amsmath} 
\usepackage{amssymb}
\usepackage{amsfonts}
\usepackage{upgreek} 

\newcommand*\patchAmsMathEnvironmentForLineno[1]{%
\expandafter\let\csname old#1\expandafter\endcsname\csname #1\endcsname
\expandafter\let\csname oldend#1\expandafter\endcsname\csname
end#1\endcsname
 \renewenvironment{#1}%
   {\linenomath\csname old#1\endcsname}%
   {\csname oldend#1\endcsname\endlinenomath}%
}
\newcommand*\patchBothAmsMathEnvironmentsForLineno[1]{%
  \patchAmsMathEnvironmentForLineno{#1}%
  \patchAmsMathEnvironmentForLineno{#1*}%
}
\AtBeginDocument{%
\patchBothAmsMathEnvironmentsForLineno{equation}%
\patchBothAmsMathEnvironmentsForLineno{align}%
\patchBothAmsMathEnvironmentsForLineno{flalign}%
\patchBothAmsMathEnvironmentsForLineno{alignat}%
\patchBothAmsMathEnvironmentsForLineno{gather}%
\patchBothAmsMathEnvironmentsForLineno{multline}%
\patchBothAmsMathEnvironmentsForLineno{eqnarray}%
}

\usepackage{hyperxmp}

\usepackage[pdftex,
            pdfauthor={\paperauthors},
            pdftitle={\paperasciititle},
            pdfkeywords={\paperkeywords},
            pdfcopyright={Copyright (C) \papercopyright},
            pdflicenseurl={\paperlicenceurl}]{hyperref}

\usepackage[all]{hypcap} 

\usepackage{xspace} 
\usepackage{upgreek}

\def\lhcb {\mbox{LHCb}\xspace}

\def\MagUp {\mbox{\em Mag\kern -0.05em Up}\xspace}

\ifthenelse{\boolean{uprightparticles}}%
{

 \def\Ppi         {\ensuremath{\uppi}\xspace}

 \def\Ppsi        {\ensuremath{\uppsi}\xspace}

 \def\PDelta      {\ensuremath{\Delta}\xspace}                 
 \def\PXi      {\ensuremath{\Xi}\xspace}                 
 \def\PLambda      {\ensuremath{\Lambda}\xspace}                 
 \def\PSigma      {\ensuremath{\Sigma}\xspace}                 
 \def\POmega      {\ensuremath{\Omega}\xspace}                 
 \def\PUpsilon      {\ensuremath{\Upsilon}\xspace}                 
 
 \def\PB      {\ensuremath{\mathrm{B}}\xspace}                 
                  
 \def\PD      {\ensuremath{\mathrm{D}}\xspace}

 \def\PJ      {\ensuremath{\mathrm{J}}\xspace}                 
 \def\PK      {\ensuremath{\mathrm{K}}\xspace}

 \def\PZ      {\ensuremath{\mathrm{Z}}\xspace}                 
                  
 \def\Pb      {\ensuremath{\mathrm{b}}\xspace}                 
 \def\Pc      {\ensuremath{\mathrm{c}}\xspace}

 \def\Ph      {\ensuremath{\mathrm{h}}\xspace}                 
 \def\Pi      {\ensuremath{\mathrm{i}}\xspace}

 \def\Ps      {\ensuremath{\mathrm{s}}\xspace}

}
{

 \def\Ppi         {\ensuremath{\pi}\xspace}

 \def\Ppsi        {\ensuremath{\psi}\xspace}                 
                  
 \mathchardef\PDelta="7101
 \mathchardef\PXi="7104
 \mathchardef\PLambda="7103
 \mathchardef\PSigma="7106
 \mathchardef\POmega="710A
 \mathchardef\PUpsilon="7107
                  
 \def\PB      {\ensuremath{B}\xspace}                 
                  
 \def\PD      {\ensuremath{D}\xspace}

 \def\PJ      {\ensuremath{J}\xspace}                 
 \def\PK      {\ensuremath{K}\xspace}

 \def\PZ      {\ensuremath{Z}\xspace}                 
                  
 \def\Pb      {\ensuremath{b}\xspace}                 
 \def\Pc      {\ensuremath{c}\xspace}

 \def\Ph      {\ensuremath{h}\xspace}                 
 \def\Pi      {\ensuremath{i}\xspace}

 \def\Ps      {\ensuremath{s}\xspace}

}

\makeatletter
\ifcase \@ptsize \relax
  \newcommand{\miniscule}{\@setfontsize\miniscule{4}{5}}
\or
  \newcommand{\miniscule}{\@setfontsize\miniscule{5}{6}}
\or
  \newcommand{\miniscule}{\@setfontsize\miniscule{5}{6}}
\fi
\makeatother

\DeclareRobustCommand{\optbar}[1]{\shortstack{{\miniscule (\rule[.5ex]{1.25em}{.18mm})}
  \\ [-.7ex] $#1$}}

\def\Z      {{\ensuremath{\PZ}}\xspace}

\def\squark    {{\ensuremath{\Ps}}\xspace}

\def\cquark    {{\ensuremath{\Pc}}\xspace}

\def\bquark    {{\ensuremath{\Pb}}\xspace}

\def\pion   {{\ensuremath{\Ppi}}\xspace}

\def\pip    {{\ensuremath{\pion^+}}\xspace}
\def\pim    {{\ensuremath{\pion^-}}\xspace}

\def\kaon    {{\ensuremath{\PK}}\xspace}
  \def\Kbar    {{\kern 0.2em\overline{\kern -0.2em \PK}{}}\xspace}

\def\KorKbar    {\kern 0.18em\optbar{\kern -0.18em K}{}\xspace}

\def\Kp      {{\ensuremath{\kaon^+}}\xspace}

  \def\Dbar    {{\kern 0.2em\overline{\kern -0.2em \PD}{}}\xspace}

\def\DorDbar    {\kern 0.18em\optbar{\kern -0.18em D}{}\xspace}

\def\B       {{\ensuremath{\PB}}\xspace}
\def\Bbar    {{\ensuremath{\kern 0.18em\overline{\kern -0.18em \PB}{}}}\xspace}

\def\BorBbar    {\kern 0.18em\optbar{\kern -0.18em B}{}\xspace}
\def\Bz      {{\ensuremath{\B^0}}\xspace}

\def\Bu      {{\ensuremath{\B^+}}\xspace}

\def\Bp      {{\ensuremath{\Bu}}\xspace}

\def\Bd      {{\ensuremath{\B^0}}\xspace}
\def\Bs      {{\ensuremath{\B^0_\squark}}\xspace}

\def\jpsi     {{\ensuremath{{\PJ\mskip -3mu/\mskip -2mu\Ppsi\mskip 2mu}}}\xspace}

  \def\Y#1S{\ensuremath{\PUpsilon{(#1S)}}\xspace}

\def\Lz          {{\ensuremath{\PLambda}}\xspace}
\def\Lbar        {{\ensuremath{\kern 0.1em\overline{\kern -0.1em\PLambda}}}\xspace}
\def\LorLbar    {\kern 0.18em\optbar{\kern -0.18em \PLambda}{}\xspace}

\def\Lb      {{\ensuremath{\Lz^0_\bquark}}\xspace}

\def\BF         {{\ensuremath{\mathcal{B}}}\xspace}

\def\BR         {\BF}
\newcommand{\decay}[2]{\ensuremath{#1\!\to #2}\xspace}         

\def\to                 {\ensuremath{\rightarrow}\xspace}

\def\BTohh        {\decay{\B}{\Ph^+ \Ph'^-}}

\def\BdToKpi      {\decay{\Bd}{\Kp\pim}}

\def\AT#1     {\ensuremath{A_{\mathrm{T}}^{#1}}\xspace}           

\def\C#1      {\ensuremath{\mathcal{C}_{#1}}\xspace}                       
\def\Cp#1     {\ensuremath{\mathcal{C}_{#1}^{'}}\xspace}                    
\def\Ceff#1   {\ensuremath{\mathcal{C}_{#1}^{\mathrm{(eff)}}}\xspace}        
\def\Cpeff#1  {\ensuremath{\mathcal{C}_{#1}^{'\mathrm{(eff)}}}\xspace}       
\def\Ope#1    {\ensuremath{\mathcal{O}_{#1}}\xspace}                       
\def\Opep#1   {\ensuremath{\mathcal{O}_{#1}^{'}}\xspace}                    

\newcommand{\tev}{\ifthenelse{\boolean{inbibliography}}{\ensuremath{~T\kern -0.05em eV}}{\ensuremath{\mathrm{\,Te\kern -0.1em V}}}\xspace}
\newcommand{\gev}{\ensuremath{\mathrm{\,Ge\kern -0.1em V}}\xspace}
\newcommand{\mev}{\ensuremath{\mathrm{\,Me\kern -0.1em V}}\xspace}
\newcommand{\kev}{\ensuremath{\mathrm{\,ke\kern -0.1em V}}\xspace}
\newcommand{\ev}{\ensuremath{\mathrm{\,e\kern -0.1em V}}\xspace}
\newcommand{\gevc}{\ensuremath{{\mathrm{\,Ge\kern -0.1em V\!/}c}}\xspace}
\newcommand{\mevc}{\ensuremath{{\mathrm{\,Me\kern -0.1em V\!/}c}}\xspace}
\newcommand{\gevcc}{\ensuremath{{\mathrm{\,Ge\kern -0.1em V\!/}c^2}}\xspace}
\newcommand{\gevgevcccc}{\ensuremath{{\mathrm{\,Ge\kern -0.1em V^2\!/}c^4}}\xspace}
\newcommand{\mevcc}{\ensuremath{{\mathrm{\,Me\kern -0.1em V\!/}c^2}}\xspace}

\def\mum  {\ensuremath{{\,\upmu\mathrm{m}}}\xspace}

\def\invfb   {\ensuremath{\mbox{\,fb}^{-1}}\xspace}

\newcommand{\stat}{\ensuremath{\mathrm{\,(stat)}}\xspace}
\newcommand{\syst}{\ensuremath{\mathrm{\,(syst)}}\xspace}

\newcommand{\chisq}{\ensuremath{\chi^2}\xspace}

\newcommand{\chisqip}{\ensuremath{\chi^2_{\text{IP}}}\xspace}

\def\gsim{{~\raise.15em\hbox{$>$}\kern-.85em
          \lower.35em\hbox{$\sim$}~}\xspace}
\def\lsim{{~\raise.15em\hbox{$<$}\kern-.85em
          \lower.35em\hbox{$\sim$}~}\xspace}

\def\ptot       {\mbox{$p$}\xspace}
\def\pt         {\mbox{$p_{\mathrm{ T}}$}\xspace}
\def\et         {\mbox{$E_{\mathrm{ T}}$}\xspace}

\def\evtgen     {\mbox{\textsc{EvtGen}}\xspace}

\def\geant      {\mbox{\textsc{Geant4}}\xspace}

\def\photos     {\mbox{\textsc{Photos}}\xspace}

\def\pythia     {\mbox{\textsc{Pythia}}\xspace}

\def\tell1  {TELL1\xspace}
\def\ukl1   {UKL1\xspace}

\newcommand{\eg}{\mbox{\itshape e.g.}\xspace}

\def\memu    {\ensuremath{m_{e^{\pm}\mu^{\mp}}}\xspace}
\def\Bds     {{\ensuremath{\B^0_{(\squark)}}}\xspace}
\def\Bsmumu    {\decay{\Bs}{\mu^+\mu^-}}
\def\Bsemu    {\decay{\Bs}{e^\pm\mu^\mp}}
\def\Bdemu    {\decay{\Bd}{e^\pm\mu^\mp}}
\def\Bdsemu    {\decay{\Bds}{e^\pm\mu^\mp}}

\def\BuToJpsiK  {\decay{\Bu}{\jpsi\Kp}}
\def\BuToJPsiK  {\BuToJpsiK}

\def\BuJpsiK {\BuToJpsiK}
\def\BdKpi  {\BdToKpi}
\def\Jee    {\decay{\jpsi}{e^+e^-}}
\def\Jmm    {\decay{\jpsi}{\mu^+\mu^-}}

\def\CLs {\ensuremath{\mathrm{CL_s}}\xspace}

\usepackage{cite} 
\usepackage{mciteplus}
\usepackage{placeins}
\usepackage{hyperref}

\usepackage{longtable} 

\begin{document}

\renewcommand{\thefootnote}{\fnsymbol{footnote}}
\setcounter{footnote}{1}

\begin{titlepage}
\pagenumbering{roman}

\vspace*{-1.5cm}
\centerline{\large EUROPEAN ORGANIZATION FOR NUCLEAR RESEARCH (CERN)}
\vspace*{1.5cm}
\noindent
\begin{tabular*}{\linewidth}{lc@{\extracolsep{\fill}}r@{\extracolsep{0pt}}}
\ifthenelse{\boolean{pdflatex}}
{\vspace*{-2.7cm}\mbox{\!\!\!\includegraphics[width=.14\textwidth]{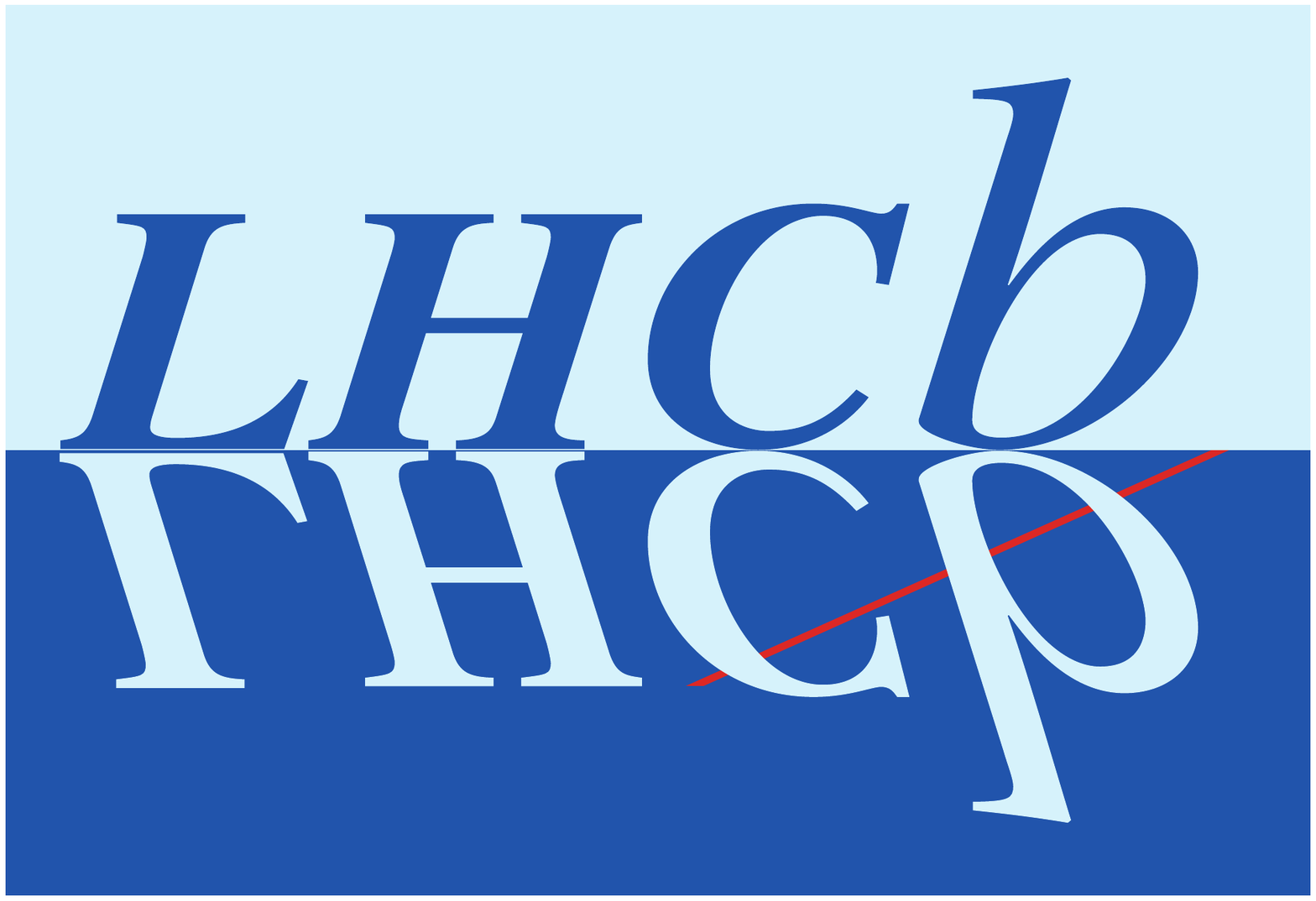}} & &}%
{\vspace*{-1.2cm}\mbox{\!\!\!\includegraphics[width=.12\textwidth]{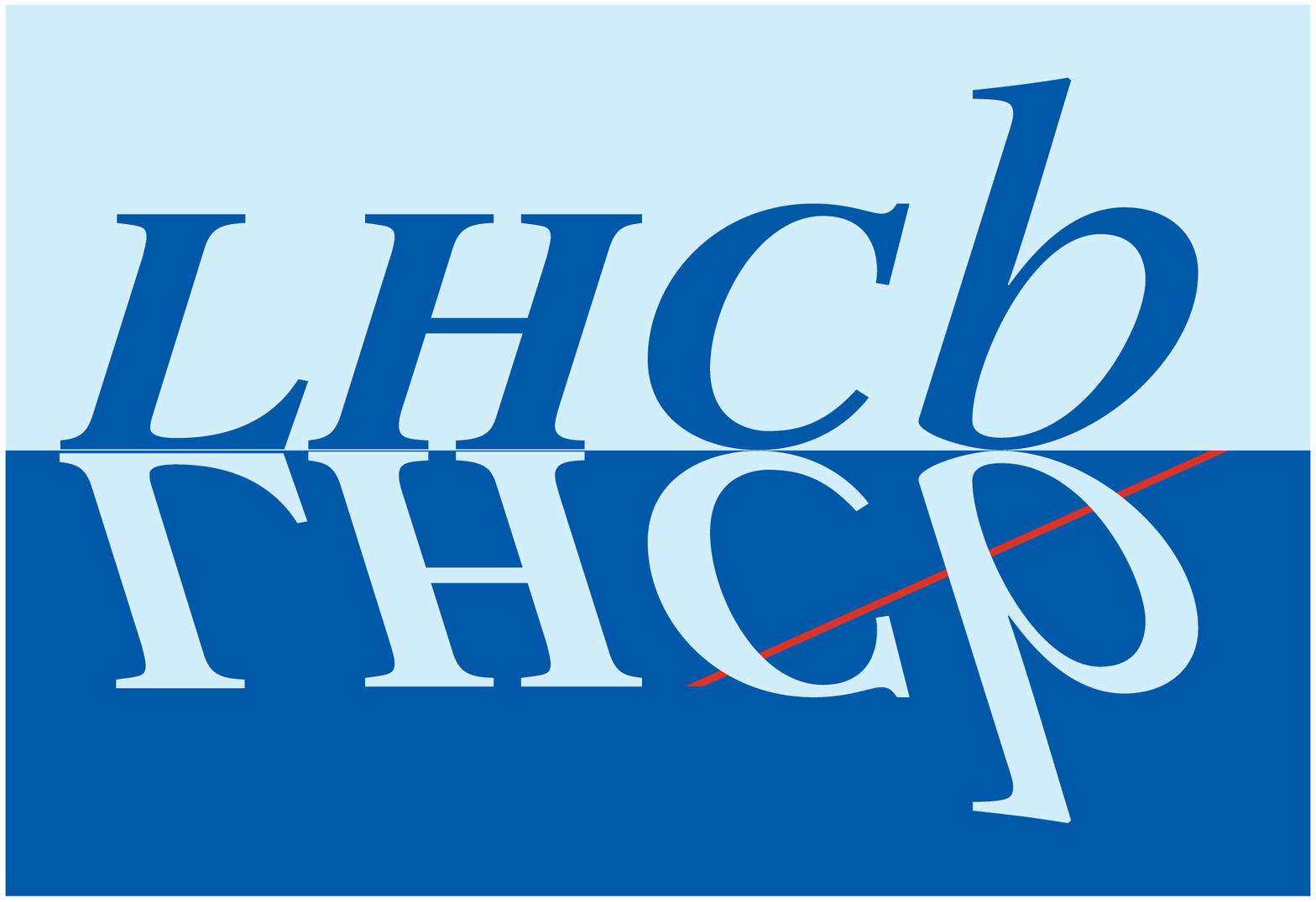}} & &}%
\\
 & & CERN-EP-2017-242 \\  
 & & LHCb-PAPER-2017-031 \\
 & & 11 October 2017 \\ 
 & & \\

\end{tabular*}

\vspace*{4.0cm}

{\normalfont\bfseries\boldmath\huge
\begin{center}

  \papertitle 
\end{center}
}

\vspace*{2.0cm}

\begin{center}
\paperauthors\footnote{Authors are listed at the end of this paper.}
\end{center}

\vspace{\fill}

\begin{abstract}
  \noindent
  A search for the lepton-flavour violating decays \Bsemu and \Bdemu is performed based on a sample of proton-proton collision data corresponding to an integrated luminosity of 3\invfb, collected with the LHCb experiment at centre-of-mass energies of 7 and 8\tev. The observed yields are consistent with the background-only hypothesis. Upper limits on the branching fraction of the \Bsemu decays are evaluated both in the hypotheses of an amplitude completely dominated by the heavy eigenstate and by the light eigenstate. The results are \mbox{$\BR(\Bsemu) < 6.3\,(5.4) \times 10^{-9}$} and \mbox{$\BF(\Bsemu) < 7.2\,(6.0) \times 10^{-9}$} at $95\%\,(90\%)$ confidence level, respectively. The upper limit on the branching fraction of the \Bdemu decay is also evaluated, obtaining \mbox{$\BR(\Bdemu) < 1.3\,(1.0) \times 10^{-9}$} at $95\%\,(90\%)$ confidence level. These are the strongest limits on these decays to date.
  
\end{abstract}

\vspace*{2.0cm}

\begin{center}
  Published in JHEP 03 (2018) 078 
\end{center}

\vspace{\fill}

{\footnotesize 
\centerline{\copyright~\papercopyright, licence \href{\paperlicenceurl}{\paperlicence}.}}
\vspace*{2mm}

\end{titlepage}


\newpage
\setcounter{page}{2}
\mbox{~}

\cleardoublepage

\renewcommand{\thefootnote}{\arabic{footnote}}
\setcounter{footnote}{0}

\pagestyle{plain}
\setcounter{page}{1}
\pagenumbering{arabic}


\section{Introduction}
\label{sec:Introduction}

Processes that are suppressed or forbidden in the Standard Model (SM) are sensitive to potential 
contributions from new mediators, even if their masses are inaccessible to direct searches.
Despite the fact that lepton-flavour violating (LFV) decays are forbidden within the SM, 
neutrino oscillation phenomena are proof that lepton flavour is not conserved in the neutral sector. 
However,
LFV decays have not yet been observed, and their observation would be clear evidence of physics beyond the SM.

The study of LFV decays is particularly interesting in light
of hints of lepton non-universality (LNU) effects in semileptonic 
decays~\cite{LHCb-PAPER-2015-025} and $b\to s\ell\ell$
transitions~\cite{LHCb-PAPER-2017-013,LHCb-PAPER-2014-024},
which could be associated with LFV processes~\cite{Guadagnoli}. 
Possible explanations of these hints can be found in various scenarios beyond the SM,
\eg models with a new gauge $\Z^\prime$ boson~\cite{Crivellin} or leptoquarks~\cite{Becirevic,Varzielas}.
In these models, the branching fractions of the \mbox{\Bsemu} and \mbox{\Bdemu} decays\footnote{Inclusion of charge conjugate processes is implied throughout the text.} can be enhanced up to $10^{-11}$. 
Other models also predict possible enhancement for \mbox{\Bsemu} and \mbox{\Bdemu} decays, \eg
heavy singlet Dirac neutrinos~\cite{ilakovac}, supersymmetric models~\cite{susy} and the Pati-Salam model\cite{patisalam}.
The most stringent published limits on the branching fractions of these decays are currently 
$\BF(\Bsemu)< 1.4 \times 10^{-8}$ and $\BF(\Bdemu) < 3.7 \times 10^{-9}$ at 95\% confidence level (CL)
from the LHCb collaboration using data corresponding to 1\invfb of integrated luminosity~\cite{LHCB-PAPER-2013-030}.

This article presents an analysis performed on a larger data sample, corresponding to an integrated luminosity
of \mbox{3\invfb} of $pp$ collisions collected at centre-of-mass energies of 7 and 8~\tev by the
LHCb experiment in 2011 and 2012.
In addition to a larger data sample, this analysis benefits from an improved selection 
and in particular a better performing multivariate classifier for signal and background separation. It supersedes 
the previous LHCb search for \Bsemu and \Bdemu decays~\cite{LHCB-PAPER-2013-030}.

Two normalisation channels are used: the \BdKpi decay which has a similar topology to that of the signal, and the \BuToJPsiK decay, with \Jmm, which has an abundant yield and a similar purity and trigger selection.
To avoid potential biases, \Bdsemu candidates in the signal region, 
$\memu \in [5100, 5500]\mevcc$, where \memu is the invariant mass of the $e^{\pm}\mu^{\mp}$ pair, were not examined until the selection and fitting procedure were finalised.

\section{Detector and simulation}
\label{sec:Detector}

The \lhcb detector~\cite{Alves:2008zz,LHCb-DP-2014-002} is a single-arm forward
spectrometer covering the \mbox{pseudorapidity} range $2<\eta <5$,
designed for the study of particles containing \bquark or \cquark quarks. 
The detector includes a high-precision tracking system
consisting of a silicon-strip vertex detector surrounding the $pp$
interaction region, a large-area silicon-strip detector located
upstream of a dipole magnet with a bending power of about
$4{\mathrm{\,Tm}}$, and three stations of silicon-strip detectors and straw
drift tubes placed downstream of the magnet.
The tracking system provides a measurement of momentum, \ptot, of charged particles with
a relative uncertainty that varies from 0.5\% at low momentum to 1.0\% at 200\gevc.
The minimum distance of a track to a primary vertex (PV), the impact parameter (IP), 
is measured with a resolution of $(15+29/\pt)\mum$,
where \pt is the component of the momentum transverse to the beam, in\,\gevc.
Different types of charged hadrons are distinguished using information
from two ring-imaging Cherenkov detectors. 
Photons, electrons and hadrons are identified by a calorimeter system consisting of
scintillating-pad and preshower detectors, an electromagnetic
calorimeter and a hadronic calorimeter. Muons are identified by a
system composed of alternating layers of iron and multiwire
proportional chambers.

The online event selection is performed by a trigger, 
which consists of a hardware stage, based on information from the muon and calorimeter
systems, followed by a software stage that applies a full reconstruction of the event.
The \Bdsemu candidates must fulfill the requirements of the electron or muon triggers.
At the hardware stage, the electron trigger requires the presence of a cluster in the electromagnetic calorimeter with a transverse energy deposit, \et,  
of at least 2.5 (3.0)\gev for 2011 (2012) data. The muon trigger selects muon candidates 
with \pt higher than $1.5$ ($1.8$)\gevc for 2011 (2012) data.
The software stage requires a two-track secondary vertex identified by a multivariate algorithm~\cite{Gligorov:2012qt} to be consistent 
with the decay of a \bquark hadron with at least two charged tracks, and at least one track with high \pt and large IP with respect to any PV.

Simulated samples are used to evaluate geometrical, reconstruction and selection efficiencies for both signal and backgrounds, 
to train multivariate classifiers and to determine the shapes of invariant mass distributions of both signal and backgrounds.
In the simulation, $pp$ collisions are generated using
\pythia~\cite{Sjostrand:2007gs}
with a specific \lhcb
configuration~\cite{LHCb-PROC-2010-056}.  Decays of hadronic particles
are described by \evtgen~\cite{Lange:2001uf}, in which final-state
radiation is generated using \photos~\cite{Golonka:2005pn}. The
interaction of the generated particles with the detector, and its response,
are simulated using the \geant
toolkit~\cite{Allison:2006ve, *Agostinelli:2002hh} as described in
Ref.~\cite{LHCb-PROC-2011-006}.

\section{Selection}
\label{sec:selection}

The \Bdsemu candidates in the events passing the trigger selection are constructed by combining pairs of tracks producing good quality secondary vertices that are separated from any PV in the downstream direction by a flight distance greater than 15 times its uncertainty.
Only \Bds candidates with $\pt > 0.5$\gevc and 
a small impact parameter \chisq, \chisqip, are considered, where the \chisqip of a \Bds candidate
is defined as the difference between the \chisq of the PV reconstructed with and without
the considered candidate. The PV with the smallest \chisqip is associated to the \Bds candidate.
The measured momentum of electron candidates is corrected for the loss of momentum due to bremsstrahlung. 
This correction is made by adding to the electron the momentum of photons consistent with being 
emitted from the electron before the magnet~\cite{Kstee}.
Since bremsstrahlung can affect the kinematic distribution of \Bdsemu candidates,
the sample is split into two categories: candidates in which no photon is associated with 
the electron and candidates for which one or more photons are recovered. The fraction of electrons with recovered bremsstrahlung photons is about 60\% for \Bdsemu decays.
Only \Bdsemu candidates with $\memu \in [4900,5850] \mevcc$ are retained to be further analysed.

Particles forming the \Bdsemu candidates are required to be well identified as an electron and a muon~\cite{muonid},
using information from the Cherenkov detectors, the calorimeters and the muon stations.
These identification criteria are optimised to keep high signal efficiency while
maximising the rejection power for the two-body hadronic $B$ decays, \BTohh, which are 
the major peaking backgrounds.

In order to reduce combinatorial background --- combinations of two random tracks that can be associated to a common vertex --- a loose requirement 
on the response of a multivariate classifier trained on simulated events is applied to the signal candidates.  
This classifier takes the following geometrical variables as input: the direction of the \Bds meson candidate; 
its impact parameter with respect to the assigned PV, defined as the PV with which it forms the 
smallest \chisqip; the separation between the two outgoing leptonic tracks at their point of closest approach; and the minimum IP of each lepton particle with respect to any PV.
In total 22~020 \Bdsemu candidates are selected, which are mainly comprised of combinatorial background that is made up of true electrons and muons.

The normalisation channels are selected with requirements as similar as possible
to those used for the signal. The selection for \BdKpi candidates is the same as for the \Bdsemu  
channel, except for the particle identification criteria which are changed into hadronic particle identification requirements.
Similarly, the \BuToJPsiK candidate selection is also kept as similar as possible, applying the same selection used 
for the signal to the dimuon pair from the \jpsi, except for the particle identification requirements. 
Additionally, loose quality requirements are applied on the \Bu vertex 
and particle identification is required on both muons. 
Finally, a $60\mevcc$ mass window around the nominal \jpsi mass and the requirement $1.4 < 1 + p_{\jpsi}/p_K < 20.0$ is used. 
The latter removes backgrounds that have a least one track that is misidentified and another that is not reconstructed, mainly $\B\to\jpsi\pip X$, 
where X can be one or more particles. 

\section{BDT training and calibration}
\label{sec:bdtcalibration}

A Boosted Decision Tree (BDT) classifier is used to separate the \Bdsemu signal from the combinatorial background. 
The BDT is trained using a simulated sample of \Bsemu events to describe the signal and a data sample of same-sign $e^{\pm}\mu^{\pm}$ candidates
to describe the combinatorial background.
The following input variables are used:
the proper decay time of the \Bds candidate; the minimum \chisqip of the two leptons with respect to the assigned PV; 
the IP of the \Bds candidate with respect to its PV; the distance of closest approach between the two lepton tracks; 
the degree of isolation of the two tracks with respect to the other tracks in the same event~\cite{LHCb-PAPER-2017-001}; 
the transverse momentum of the \Bds candidate; the cosine of the angle between the muon momentum in the \Bds candidate rest
frame and the vector perpendicular to the \Bds candidate momentum and the beam axis; 
the flight distance of the \Bds candidate with respect to its PV; the \chisq of the \Bds candidate decay vertex; the maximum 
transverse momentum of the two decay products and their difference in pseudorapidity.
 
The BDT response is transformed such that it is uniformly distributed in the range [0,1] for the signal, while 
peaking at zero for the background.
The linear correlation between the BDT response and the dilepton invariant mass is found to be around 4\%.  

\begin{figure}[!t]
\centering
\includegraphics[width=0.8\textwidth]{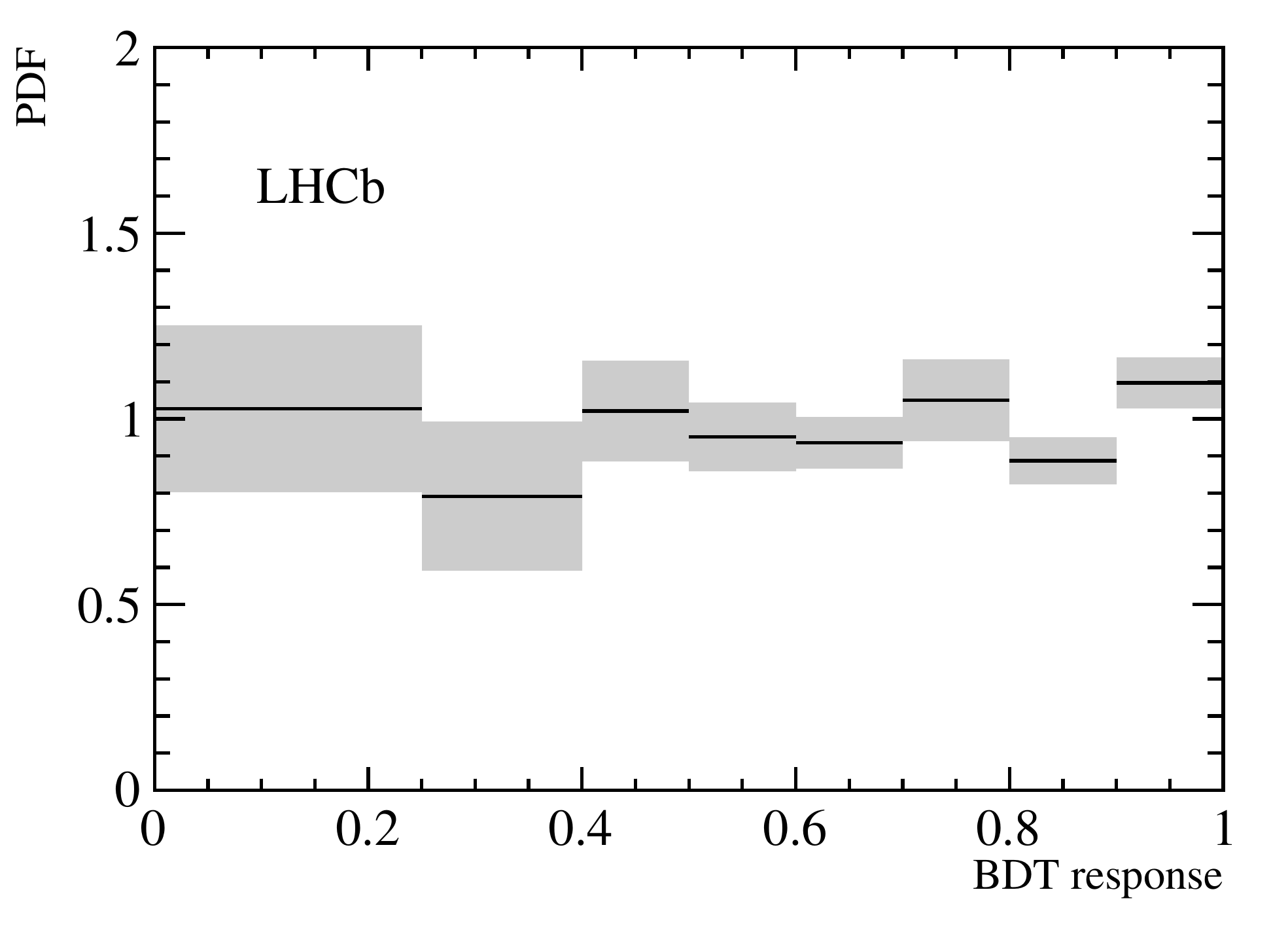}
\caption{Expected distribution of the BDT response for \Bdsemu decays with recovered bremsstrahlung photons obtained from the \BdKpi control channel.
  The total uncertainty is shown as a light grey band. Each bin is normalised to its width.} 
\label{fig:bdtresponse}
\end{figure}

Since the BDT is trained using only kinematic information of a two-body \Bds decay, 
its response is calibrated using \BdKpi decays as a proxy. 
To avoid biases, \BdKpi candidates are selected from candidates where the trigger decision did not depend 
on the presence of the \Bd decay products. 
Furthermore, the candidates are weighted to emulate the effect of the lepton triggers and the particle identification requirements.
The number of \BdKpi candidates in bins of BDT response is determined by
fitting the $\Kp\pim$ invariant mass distribution. As expected, the BDT response is found to be
consistent with a uniform distribution across the range [0,1].
The distribution of the BDT response is also checked 
on a \BdKpi simulated sample and a uniform distribution is obtained. Candidates with a value smaller than $0.25$ are then excluded, as this region is highly 
contaminated by background, leaving a total of 476 signal candidates.
The signal candidates are classified in a binned two-dimensional space formed by the BDT response and the two bremsstrahlung categories. 
The expected probability density function (PDF) of the BDT response for \Bdsemu decays with recovered bremsstrahlung photons is shown in Fig.~\ref{fig:bdtresponse}.

Unrecovered bremsstrahlung photons emitted by signal electrons can affect the BDT response and are not accounted
for in the calibration procedure since hadrons do not emit significant bremsstrahlung. 
The impact of bremsstrahlung on the BDT response distribution is evaluated using simulation and a correction 
is applied where no bremsstrahlung is recovered. 

\section{Normalisation}
\label{sec:normalisation}

\begin{figure}[!t]
\centering
\includegraphics[width=.48\textwidth]{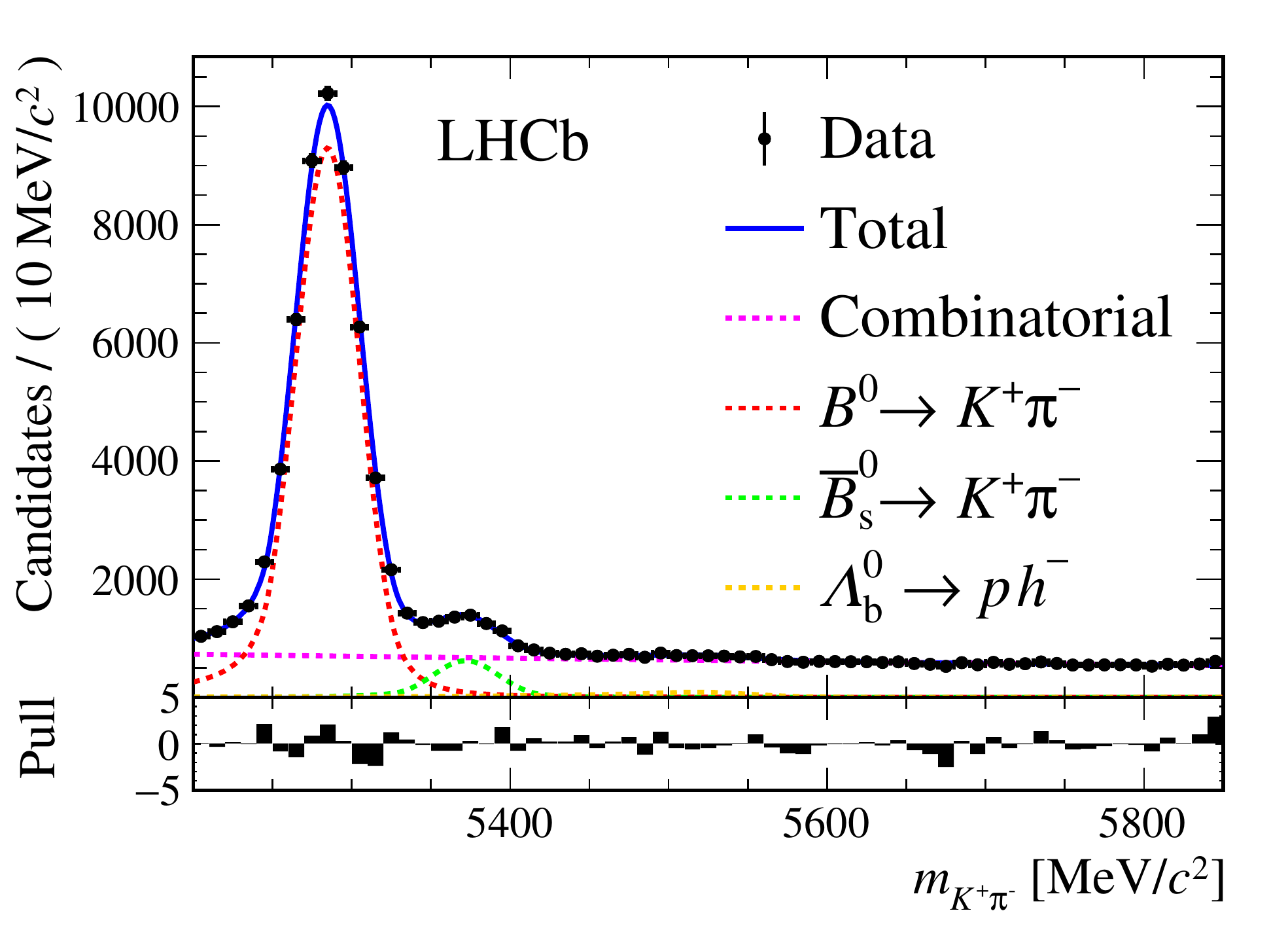}
\includegraphics[width=.48\textwidth]{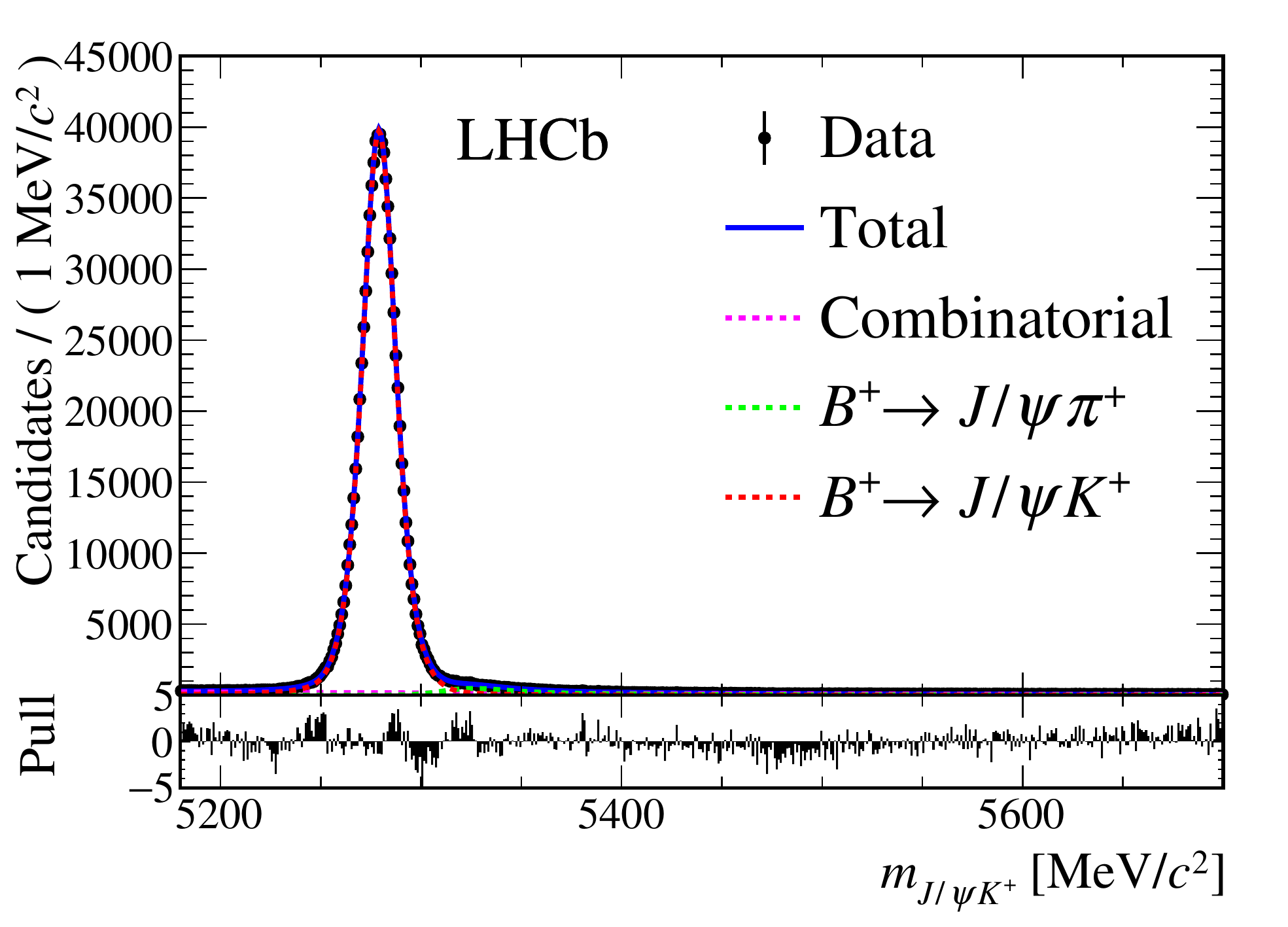}
\caption{Invariant mass distributions of the two normalisation channels with fit functions superimposed: (left) \BdKpi and (right) \BuJpsiK.
Pull distributions are shown below each plot.} 
\label{fig:normalisations}
\end{figure}

The \Bdsemu yields are obtained from a fit to the lepton-pair invariant mass distribution and translated into branching fractions according to
\begin{eqnarray}
\BF(\Bdsemu) &=& \sum_i  
    w^i \frac{{\cal B}^i_{\rm norm}}{N^i_{\rm norm}} \frac{\varepsilon^i_{\rm norm}}{\varepsilon_{\rm sig}} \frac{f_q}{f_{d(s)} } \frac{\mathcal{L}_{\rm norm}^{i}}{\mathcal{L}_{\rm sig}} 
 \times N_{\Bdsemu}  \nonumber \\
& = & \alpha_{\Bds}  \times N_{\Bdsemu},
\label{eq:normalisation}
\end{eqnarray}
where the index $i$ identifies the normalisation channel and $N^i_{\rm norm}$ and ${\cal B}^i_{\rm norm}$ are its number of candidates and its branching fraction. 
The signal yields are denoted by $N_{\Bdsemu}$ and the factors
$f_{q}$ indicate the probabilities that a \bquark quark fragments into a \Bz or \Bs meson.
Assuming $f_d=f_u$, the fragmentation probability for the $B^0$ and $B^+$ channels is set to $f_d$. 
The value of $f_s/f_d$ used is measured in $pp$ collision data at $\sqrt{s}=7$~\tev by the LHCb collaboration and is evaluated to be $0.259 \pm 0.015$~\cite{fsfd}. 
The two normalisation channels are averaged with weights $w^i$ proportional to the square of the inverse of the uncertainty related to their branching fractions and yields. A correction has also been applied for the marginal difference in luminosity, $\mathcal{L}$, between the channels.
The branching fractions of the signal decays include both charge configurations of the final-state particles, $e^+\mu^-$ and $e^-\mu^+$, 
so that {$\BF(\Bdsemu) \equiv \BF(\decay{\Bds}{e^+\mu^-}) + \BF(\decay{\Bds}{e^-\mu^+})$}.
The results of the two fits are shown in Fig.~\ref{fig:normalisations} and the measured yields are reported in Table~\ref{tab:normyields}.
 
\begin{table}[t!]
\caption{Yields of normalisation channels obtained from fits to data.}
\label{tab:normyields}
\begin{minipage}[t]{0.96\textwidth}
\begin{center}
\begin{tabular}{l|r@{}l}
\hline
& Yield\\
\hline
\BdKpi & \phantom{lll}$49\,907$&\phantom{,}$\pm$\phantom{,}$277$   \\
\BuToJPsiK & $913\,074$&\phantom{,}$\pm$\phantom{,}$1106$ \\ 
\hline
\end{tabular}
\end{center}
\end{minipage}
\end{table} 

The efficiency ${\rm \varepsilon_{sig(norm)}}$ for the signal (normalisation) channels depends on several factors:
the geometric acceptance of the detector, the probability for particles to produce hits in the detector which can be
reconstructed as tracks, and the efficiency of the selection requirements that are applied both in the trigger and selection stages, which includes 
the particle identification requirements. The ratios of acceptance, reconstruction and selection efficiencies are 
evaluated using simulation with the exception of the trigger and particle identification efficiencies, 
which are not well reproduced by simulation, and are calibrated using data~\cite{Tolk:1701134,Anderlini:2202412}.
Calibration samples where the trigger decision was independent of the candidate decay products are used to study the trigger efficiency.
From these samples, \BuJpsiK candidates, with \Jee and \Jmm, are used to study the requirements for the electrons and muons, respectively.
The efficiencies are determined as a function of the \pt and IP for the muon and \et and IP for the electron. 
The single-track efficiencies are then combined with a weighted average over the properties of
the electron and muon tracks of a \Bsemu simulated sample.
 
Particle identification efficiencies are evaluated using calibration samples where
the identity of one of the particles can be inferred by means uncorrelated to particle identification requirements. 
A tag-and-probe method is applied on {\Jmm} and {\Jee} decay samples, where only one lepton, the tag, is required to be well identified 
and the identity of the other lepton is deduced. 
The single-track efficiencies, calculated  as a function of kinematic variables, are then combined and averaged 
using the momentum distributions of the leptons in a \Bsemu simulated sample.
 
The two normalisation factors $\alpha_{\Bs}$ and $\alpha_{\Bd}$ are determined to be
$(2.48\pm0.17)\times 10^{-10}$ and $(6.16\pm0.23)\times 10^{-11}$. 
The total efficiencies for the \Bdemu, \Bsemu, \BuJpsiK and \BdKpi decays are respectively $(2.22\pm0.05)\%$, $(2.29\pm0.05)\%$, $(2.215\pm0.035)\%$ and $(0.360\pm0.021)\%$,
where the efficiencies for \Bdsemu are for the full BDT and bremsstrahlung category range.

To validate the normalisation procedure, 
the ratio between the measured branching fractions of \BdKpi and \BuJpsiK is determined as
\begin{equation}
  R_{\rm norm} = \frac{N_{\BdKpi} \times \varepsilon_{\BuJpsiK}}{ N_{\BuJpsiK} \times \varepsilon_{\BdKpi}} = 0.332 \pm 0.002\stat \pm 0.020 \syst,
\end{equation}
where $\varepsilon_{\BuJpsiK}$ and $\varepsilon_{\BdKpi}$ are the selection efficiencies for the {\BdKpi} and {\BuJpsiK} decays respectively. A correction of about 1\% is applied in order to take into account the difference in luminosity between the two channels. 
The value obtained for $R_{\rm norm}$ is in excellent agreement with the measured value of $0.321 \pm 0.013$~\cite{PDG2017}.

\section{Backgrounds}
\label{sec:backgrounds}

In addition to the combinatorial background, 
the signal region is also potentially polluted by backgrounds from exclusive decays
where one or more of the final-state particles are misidentified or not reconstructed.
The potentially most dangerous of these backgrounds are hadronic \BTohh decays where both hadrons
are misidentified as an electron-muon pair, resulting in peaking structures near the \Bsemu signal mass.
Other decays which could contribute, especially at low invariant masses, are
\mbox{$B^+_c \to \jpsi \ell'^+ \nu_{\ell'}$} with $\jpsi \rightarrow \ell^+ \ell^-$, $\Bz \to \pi^- \ell^+ \nu_{\ell}$,
$\Lb \to p \ell^- \overline{\nu}_{\ell}$ and $\Bp \to \pi^{+} \jpsi$ with $\jpsi\to\ell^+ \ell^-$,
where $\ell/\ell'^{\pm} = e^{\pm}$ or $\mu^{\pm}$. These decays do not peak under the signal
but are potentially abundant.
The expected number of candidates from each possible background decay that pass the signal selection is evaluated using simulation.
The candidates are normalised to the number of \BuToJPsiK decays found in data as
\begin{equation}
   N_{X} = N_{\BuToJPsiK} \frac{f_q}{f_u} \frac{\BR(X)}{\BR(\BuToJPsiK)\cdot\BR(\Jmm)} \frac{\varepsilon(X)}{\varepsilon(\BuToJPsiK)},
\end{equation}
where $N_{X}$ is the expected number of candidates from the $X$ decay that fall into the \Bsemu signal mass window;
$f_{q}$ is the fragmentation fraction;
$\BR(X)$, $\BR(\BuToJPsiK)$ and $\BR(\Jmm)$ are respectively the branching fractions of the decay under
study, \BuToJPsiK and \Jmm~\cite{PDG2017};
$\varepsilon(X)$ is the efficiency for each considered decay to pass the \Bsemu selection; and
$\varepsilon(\BuToJPsiK)$ is the efficiency for \BuToJPsiK candidates to pass the respective selection.
 
The mass and BDT distributions of these background modes are evaluated using simulated
samples, while the probabilities of misidentifying kaons, pions and protons as muons or electrons
are determined from $D^{*+} \to D^0 \pi^+$ with $D^0 \to K^{-} \pi^{+}$ and $\Lz \to p \pi^-$
decays selected from data.
The expected total number of \BTohh candidates is $0.11\pm0.02$ in the full BDT range,
which is negligible. This yield estimation is cross-checked using data. A sample of \BTohh decays is
selected by applying only a partial \Bdsemu selection: only the signal electron PID requirements
are applied while the second particle is required to be identified as a pion.
The application of these criteria still leaves a sizeable peak to be fit in data.
The yield of decays identified as $\Bds \to e^\pm\pi^\mp$ is then modified
to take into account the probability of a pion to be misidentified as a muon.
After this correction the expected yield is compatible with the yield obtained using the simulation.
 
The expected yields of most of the other backgrounds are also found to be negligible.
The only backgrounds which are relevant are $\Bz \to \pi^- \mu^+ \nu_{\ell}$
and $\Lb \to p \ell^- \overline{\nu}_{\ell}$ for which $55 \pm 3$ and $82 \pm 39$ candidates, respectively,
are expected in the full BDT range. The contributions from these two decays are included in the fit model.

\section{Mass calibration}
\label{sec:masscalibration}

\begin{figure}
\label{fig:bdt_calib_plots}
\centering
\includegraphics[width=.48\textwidth]{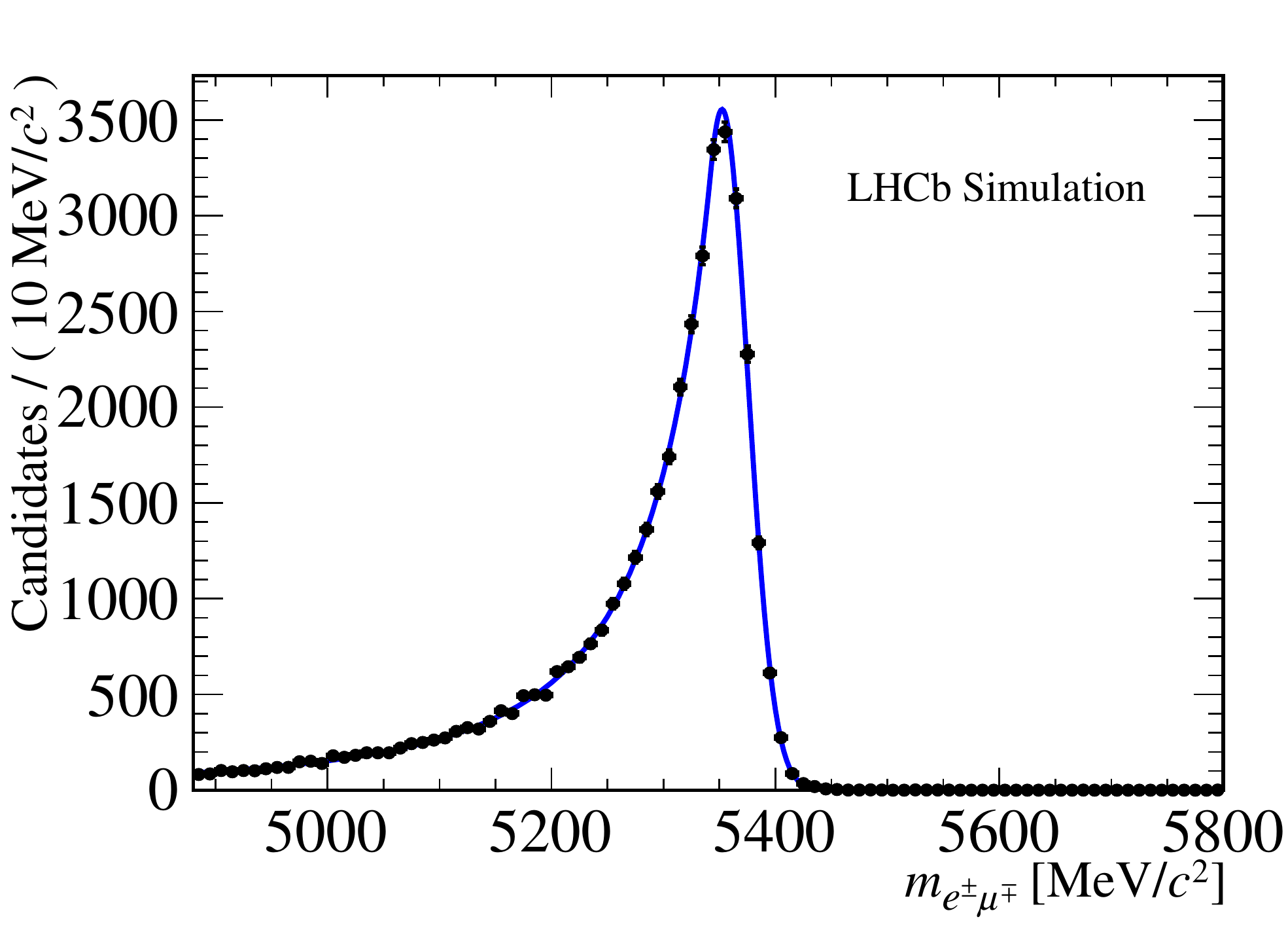}
\includegraphics[width=.48\textwidth]{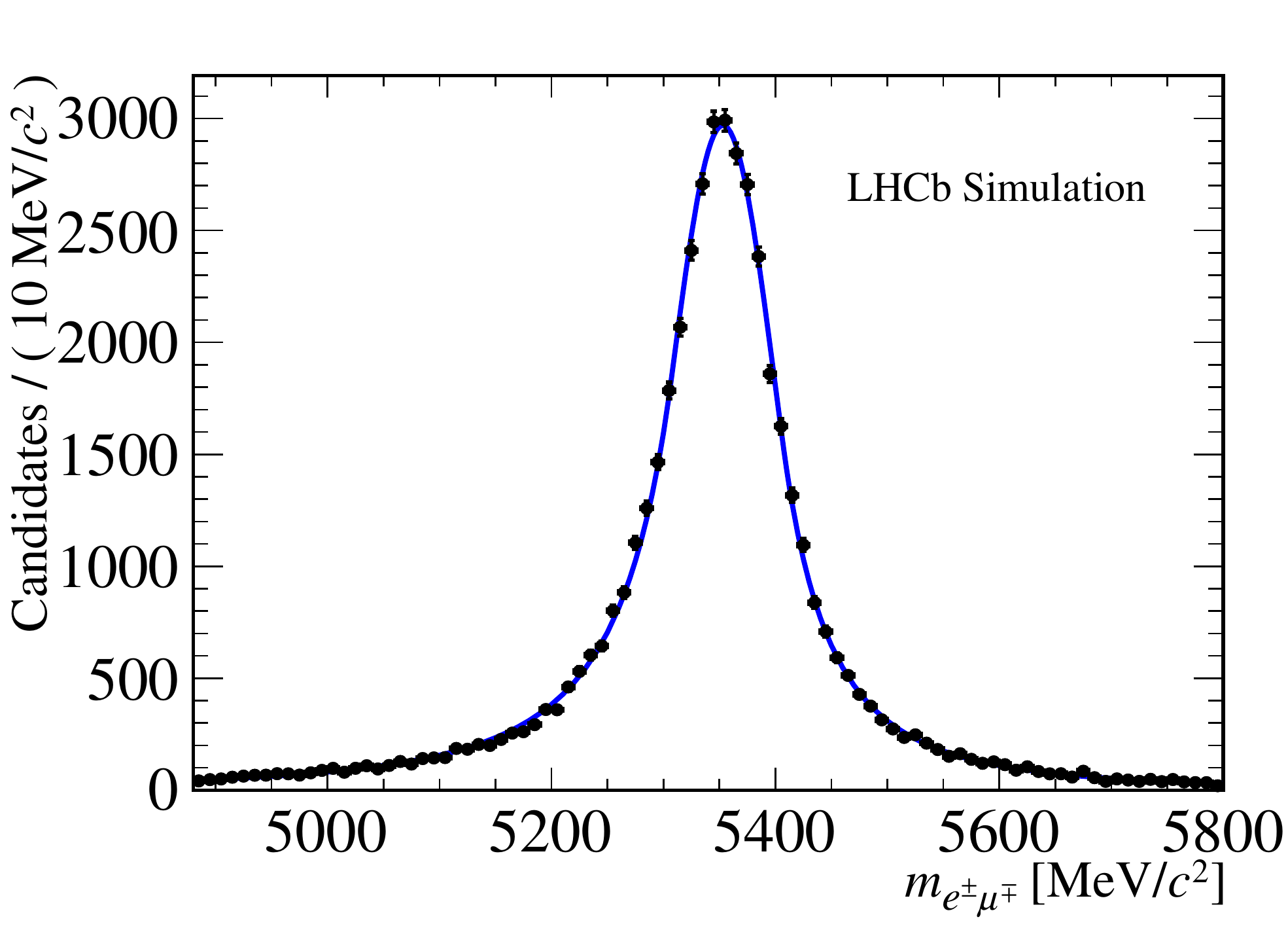}
\caption{Distribution of the \memu invariant mass of simulated \Bs candidates with no (left) and one or more (right) 
recovered bremsstrahlung photons. The overlaid fit function is a modified Crystal Ball function with two tails on opposite sides.
}
\label{fig:masscal}
\end{figure}

The invariant-mass distribution of \Bdsemu candidates is modelled
by a modified Crystal Ball function~\cite{Skwarnicki:1986xj} with two tails on opposite sides
defined by two parameters each.
The signal shape parameters are obtained from simulation, with data-driven scale factors applied
to the core resolution to correct for possible data-simulation discrepancies.
For this purpose, since there is no appropriate control channel with an electron and a muon in the
final state, \mbox{\Jee} and \mbox{\Jmm} decays are analysed comparing the
mass resolution in data and simulation. 
The results are then combined to reproduce the effect on an $e^{\pm}\mu^{\mp}$ final state. 
Corrections to the widths of the mass are of the order of 10\%.
Since bremsstrahlung can significantly alter the mass shape by enhancing the tails,
the fit model for \Bdsemu candidates is obtained separately for the two bremsstrahlung categories (see Fig.~\ref{fig:masscal}).
The mass shape parameters are found to be independent of the particular BDT bin chosen and 
a single model for each bremsstrahlung category is therefore used.

\section{Results}
\label{sec:results}

\begin{figure}[!ht]
\centering
\includegraphics[width=\textwidth]{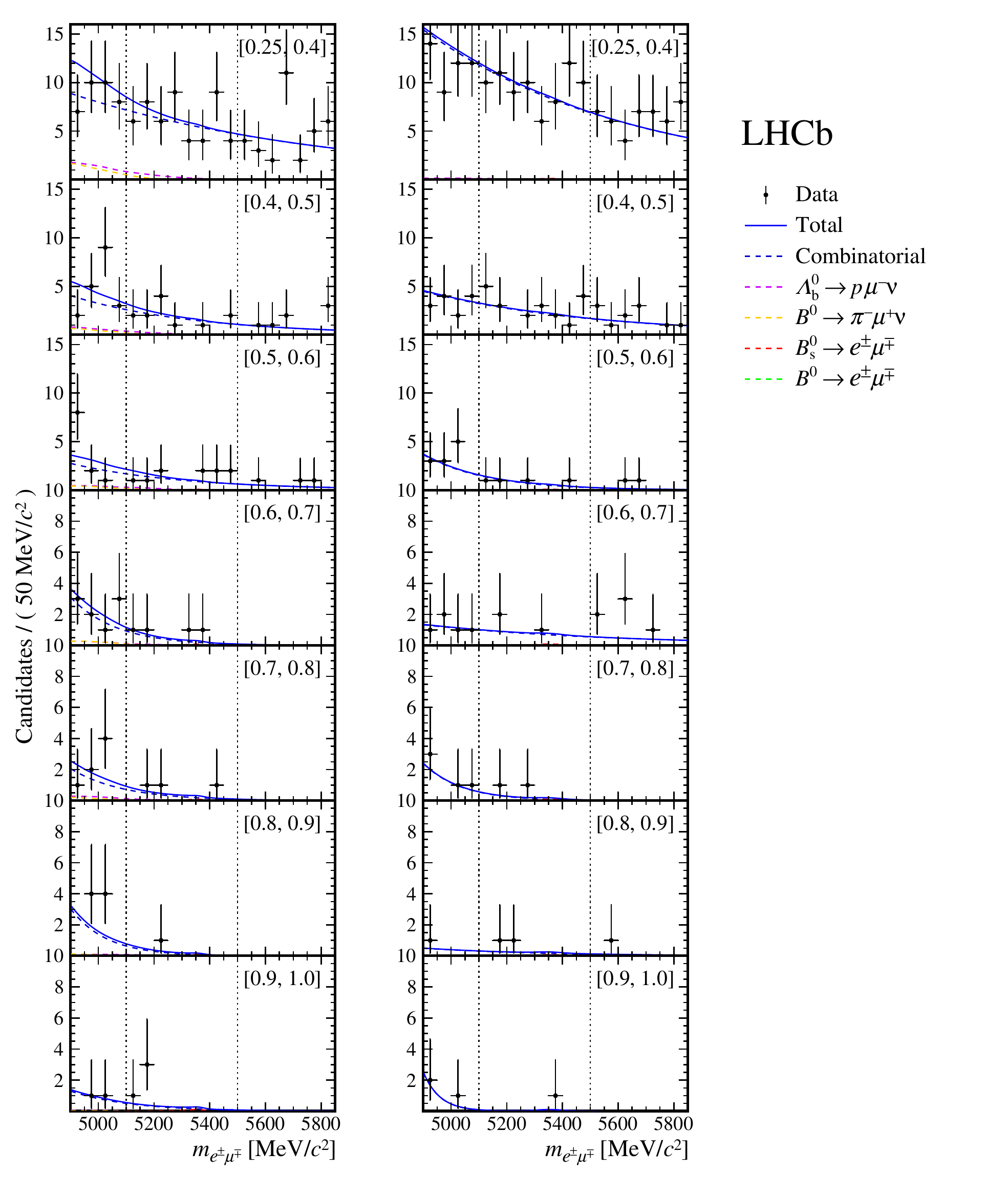}
\caption{Distributions of the invariant mass of the \Bdsemu candidates, \memu, divided into bins of BDT response and two bremsstrahlung categories (left) without and (right) 
with bremsstrahlung photons recovered. The result of the fit is overlaid and the different components are detailed.
The edges of the range that was examined only after finalising the selection and fit procedure are delimited by gray dashed vertical lines. This region includes 90\% of the potential signal candidates. Given the result obtained from the fit, 
the \Bdemu component is not visible in the plots.
}
\label{fig:fit}
\end{figure}

The data sample is split into two bremsstrahlung categories, which are further divided into seven subsets each depending on 
the BDT response covering the range from 0.25 to 1.0, with boundaries 0.25, 0.4, 0.5, 0.6, 0.7, 0.8, 0.9 and 1.0. 
The region with BDT response lower than 0.25, which is mostly populated by combinatorial background, is excluded from the fit. 
The \Bdemu and \Bsemu yields are obtained from a single unbinned extended maximum likelihood fit performed simultaneously to
the \memu distributions in each subset. The \Bdsemu fractional yields and 
the mass shape parameters in each category are Gaussian-constrained according to their expected values and uncertainties.   
The combinatorial background is modelled with an exponential function with independent yield and shape parameters in each subset.
The exclusive backgrounds are included as separate components in the fit. 
Their mass shapes are modelled using nonparametric functions determined from simulation for each bremsstrahlung category. 
The overall yields and fractions of these backgrounds are Gaussian-constrained to their expected values. 
The result of this fit is shown in Fig.~\ref{fig:fit}. 

\begin{figure}[!ht]
\centering
\includegraphics[width=.48\textwidth]{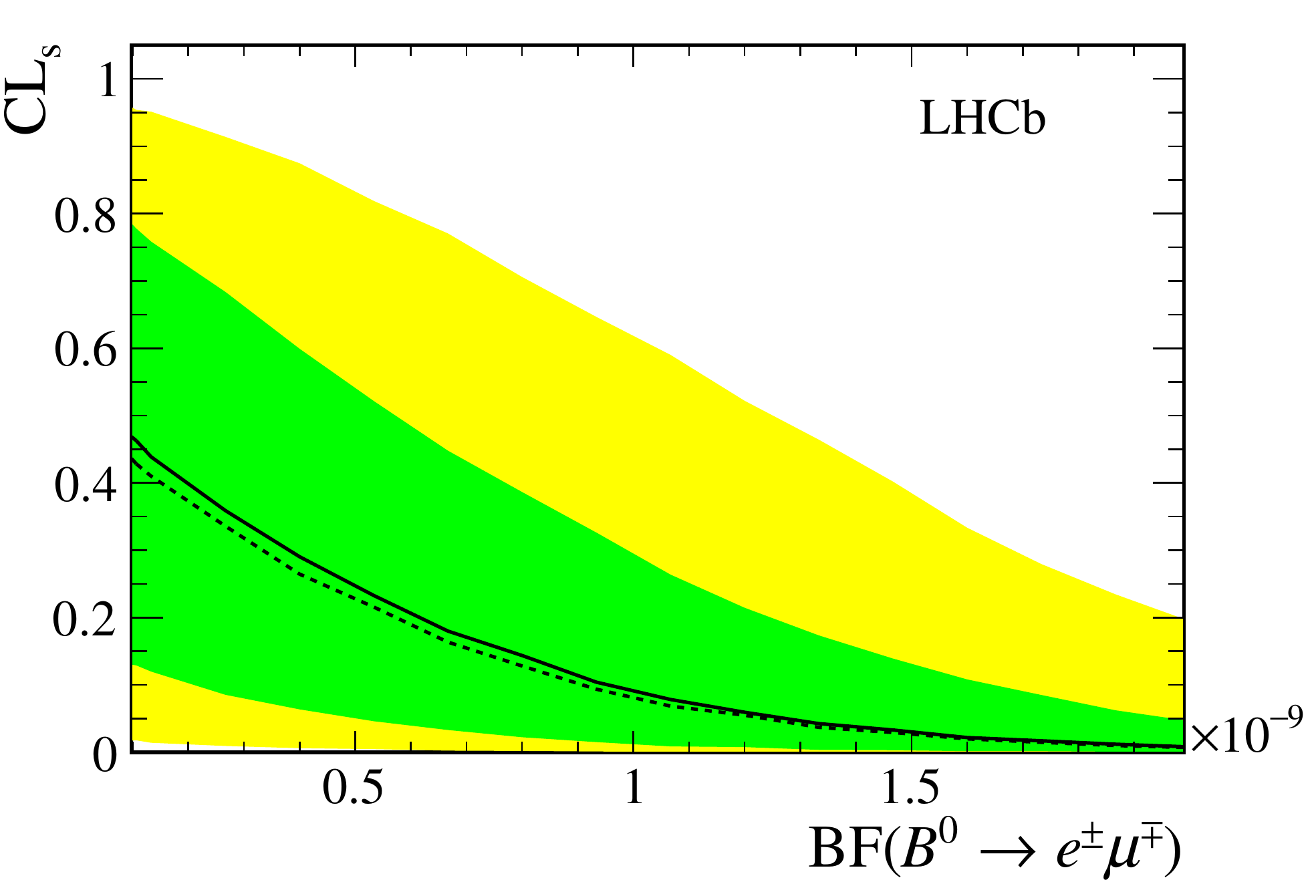}
\includegraphics[width=.48\textwidth]{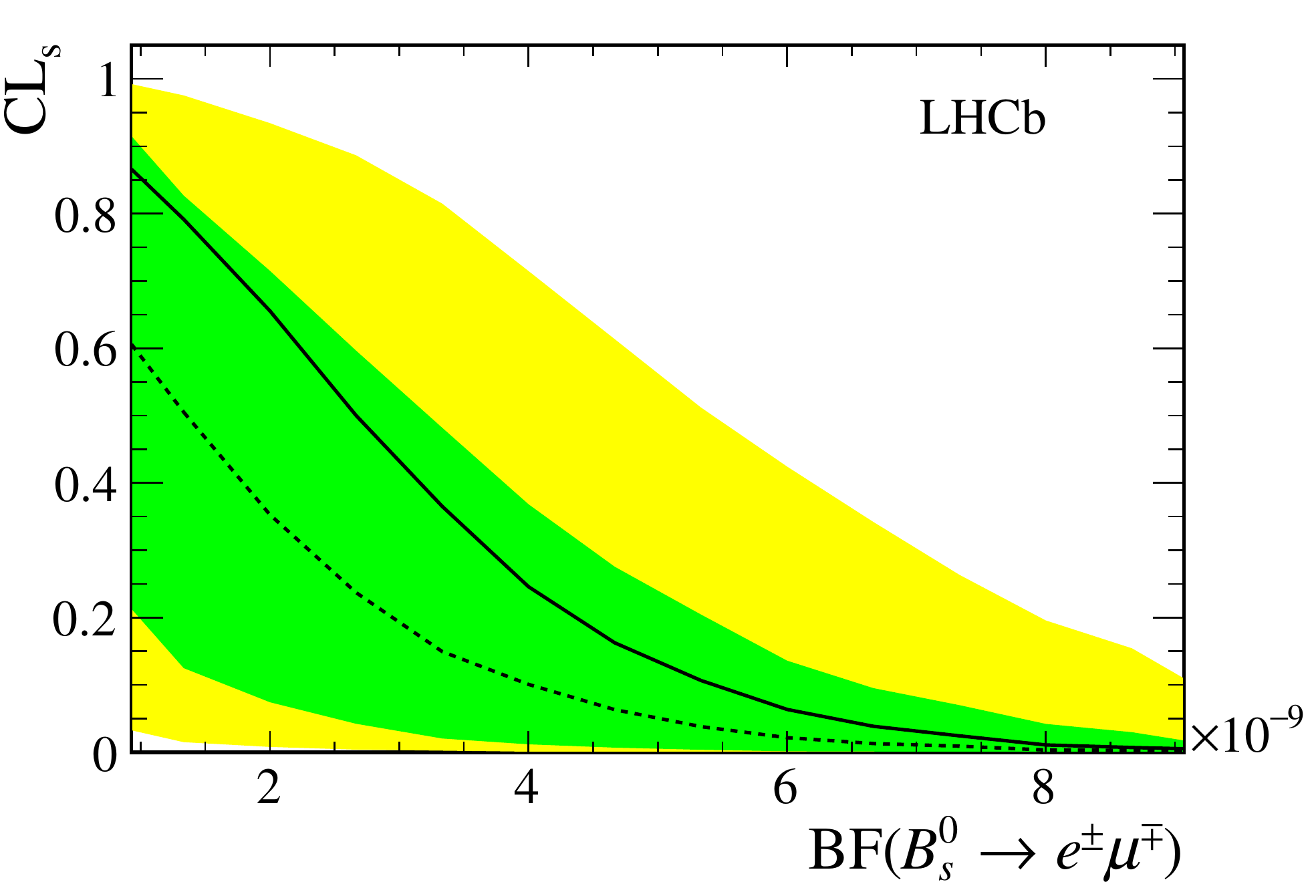}
\caption{Results of the \CLs scan used to obtain the limit on (left) $\BF(\Bdemu)$ and (right) $\BF(\Bsemu)$.
The background-only expectation is shown by the dashed line and the $1\sigma$ and $2\sigma$ bands are shown as dark (green) and light (yellow) bands respectively. 
The observed limit is shown as the solid black line.}
\label{fig:cls}
\end{figure}

No significant excess of \Bdemu or \Bsemu decays is observed and upper limits on the 
branching fractions are set using the \CLs method~\cite{Read:2002hq}. 
The ratio between the likelihoods in two hypotheses, signal plus background and background only, is used as the test statistic. 
The likelihoods are computed with nuisance parameters fixed to their nominal values. Pseudoexperiments, in which the nuisance 
parameters are varied according to their statistical and systematic uncertainties, are used for the evaluation of the test statistic. 
The resulting \CLs scans are shown in Fig.~\ref{fig:cls} and upper limits at $95\%$ and $90\%$ confidence level are reported in Table~\ref{tab:upperlimits}.

\begin{table}[t!]
\caption{Expected (assuming no signal) and observed upper limits for $\BR(\Bsemu)$ and $\BR(\Bdemu)$ at 95\%\,(90\%) CL. The upper limit on the $\BR(\Bsemu)$ 
is evaluated under the assumption of pure heavy eigenstate contribution on the decay amplitude.}
\label{tab:upperlimits}
\begin{minipage}[t]{0.96\textwidth}
\begin{center}
\begin{tabular}{l|c|c}
\hline
channel & expected & observed \\
\hline
$\BF(\Bsemu)$ & $5.0\,(3.9) \times 10^{-9}$ &  $6.3\,(5.4) \times 10^{-9}$ \\
$\BF(\Bdemu)$ &  $1.2\,(0.9) \times 10^{-9}$ &  $1.3\,(1.0) \times 10^{-9}$ \\
\hline\end{tabular}
\end{center}
\end{minipage}
\end{table}

Several systematic uncertainties can affect the evaluation of the limit on the \Bsemu and \Bdemu branching fractions
through the normalisation formula in Eq.~\ref{eq:normalisation} and the fit model used to evaluate the signal yields.
The systematic uncertainties are taken into account for the limit computation by constraining the respective nuisance parameters in the likelihood fit 
with a Gaussian distribution having the central value of the parameter as the mean and its uncertainty as the width.
The nuisance parameters for the \Bdsemu yields are related 
to the calibration of the BDT response, the parameters of the signal shape, the estimated yields of the $\Bz \to \pi^- \mu^+ \nu_{\ell}$ and 
$\Lb \to p \ell^- \overline{\nu}_{\ell}$ backgrounds and the fractional yield per bremsstrahlung category. 
For the limit on the \Bdsemu branching fractions, 
the nuisance parameters are in addition related to the signal efficiency, whose uncertainty is dominated by the systematic uncertainty on the trigger 
efficiencies, and the uncertainties on the efficiencies, branching fractions and yields of the normalisation channels. 
For the \Bsemu branching fraction estimation, Eq.~\ref{eq:normalisation} also includes the hadronisation fraction $f_s/f_d$, 
which dominates the systematic uncertainty for the normalisation. 
The overall impact on the limits is evaluated to be below 5\%.

The two \Bs mass eigenstates are characterised by a large lifetime difference. 
Depending on their contribution to the decay amplitude, the selection efficiency and the BDT shape can be affected. 
Given the negligible difference in lifetime for the \Bz system, this effect is not taken into account for the \Bdemu limit evaluation. 
Two extreme cases can be distinguished: when only the heavy or the light eigenstate contributes to the total decay amplitude. 
For example, if the only contribution to the LFV \Bsemu decay is due to neutrino oscillations, it is expected that the 
amplitude is dominated by the heavy eigenstate as for the \Bsmumu decay~\cite{LHCb-PAPER-2017-001}. 
As the contribution to the total amplitude from the heavy and light eigenstate can have an effect on the acceptance, 
the limit on $\BF(\Bsemu)$ is evaluated in the two extreme cases. The one reported in Table~\ref{tab:upperlimits} and obtained from the CLs scan in Fig.~\ref{fig:cls}, is evaluated assuming only a contribution from the heavy eigenstate. For the light eigenstate case the limit is found to be $\BF(\Bsemu) < 7.2\,(6.0) \times 10^{-9}$ at 95\%\,(90\%) CL.

\section{Summary}
\label{sec:summary}

In summary, a search for the LFV decays \Bsemu and \Bdemu is performed using $pp$ collision data collected at 
centre-of-mass energies of 7 and 8\tev, corresponding to a total integrated luminosity of 3\invfb. 
No excesses are observed for these two modes and upper limits on the branching fractions are set to $\BF(\Bsemu) < 6.3\,(5.4) \times 10^{-9}$ 
and $\BF(\Bdemu) < 1.3\,(1.0) \times 10^{-9}$ at $95\%\,(90\%)$ CL, where only a contribution from the heavy eigenstate is assumed for the $\Bs$ meson.
If the \Bs amplitude is completely dominated by the light eighenstate, the upper limit on the branching fraction becomes
$\BF(\Bsemu) < 7.2\,(6.0) \times 10^{-9}$ at $95\%\,(90\%)$ CL.
These results represent the best upper limits to date and are a factor 2 to 3 better than the previous results from LHCb~\cite{LHCB-PAPER-2013-030}.

\section*{Acknowledgements}
%
%
\noindent We express our gratitude to our colleagues in the CERN
accelerator departments for the excellent performance of the LHC. We
thank the technical and administrative staff at the LHCb
institutes. We acknowledge support from CERN and from the national
agencies: CAPES, CNPq, FAPERJ and FINEP (Brazil); MOST and NSFC
(China); CNRS/IN2P3 (France); BMBF, DFG and MPG (Germany); INFN
(Italy); NWO (The Netherlands); MNiSW and NCN (Poland); MEN/IFA
(Romania); MinES and FASO (Russia); MinECo (Spain); SNSF and SER
(Switzerland); NASU (Ukraine); STFC (United Kingdom); NSF (USA).  We
acknowledge the computing resources that are provided by CERN, IN2P3
(France), KIT and DESY (Germany), INFN (Italy), SURF (The
Netherlands), PIC (Spain), GridPP (United Kingdom), RRCKI and Yandex
LLC (Russia), CSCS (Switzerland), IFIN-HH (Romania), CBPF (Brazil),
PL-GRID (Poland) and OSC (USA). We are indebted to the communities
behind the multiple open-source software packages on which we depend.
Individual groups or members have received support from AvH Foundation
(Germany), EPLANET, Marie Sk\l{}odowska-Curie Actions and ERC
(European Union), ANR, Labex P2IO, ENIGMASS and OCEVU, and R\'{e}gion
Auvergne-Rh\^{o}ne-Alpes (France), RFBR and Yandex LLC (Russia), GVA,
XuntaGal and GENCAT (Spain), Herchel Smith Fund, the Royal Society,
the English-Speaking Union and the Leverhulme Trust (United Kingdom).

\FloatBarrier

\addcontentsline{toc}{section}{References}
\setboolean{inbibliography}{true}
\bibliographystyle{LHCb}
\bibliography{main,LHCb-PAPER,LHCb-CONF,LHCb-DP,LHCb-TDR}

\newpage


\centerline{\large\bf LHCb collaboration}
\begin{flushleft}
\small
R.~Aaij$^{40}$,
B.~Adeva$^{39}$,
M.~Adinolfi$^{48}$,
Z.~Ajaltouni$^{5}$,
S.~Akar$^{59}$,
J.~Albrecht$^{10}$,
F.~Alessio$^{40}$,
M.~Alexander$^{53}$,
A.~Alfonso~Albero$^{38}$,
S.~Ali$^{43}$
% ...

{\footnotesize \it
% ...
$ ^{5}$Clermont Universit{\'e}, Universit{\'e} Blaise Pascal, CNRS/IN2P3, LPC, Clermont-Ferrand, France\\
% ...
$ ^{10}$Fakult{\"a}t Physik, Technische Universit{\"a}t Dortmund, Dortmund, Germany\\
% ...
$ ^{38}$ICCUB, Universitat de Barcelona, Barcelona, Spain\\
$ ^{39}$Universidad de Santiago de Compostela, Santiago de Compostela, Spain\\
$ ^{40}$European Organization for Nuclear Research (CERN), Geneva, Switzerland\\
% ...
$ ^{43}$Nikhef National Institute for Subatomic Physics, Amsterdam, The Netherlands\\
% ...
$ ^{48}$H.H. Wills Physics Laboratory, University of Bristol, Bristol, United Kingdom\\
% ...
$ ^{53}$School of Physics and Astronomy, University of Glasgow, Glasgow, United Kingdom\\
% ...
$ ^{59}$University of Cincinnati, Cincinnati, OH, United States\\
% ...
}
\end{flushleft}

\end{document}